\documentclass[12pt,letterpaper]{article}
\pdfoutput=1

\usepackage{amsmath,amssymb,calc, amsthm}
\usepackage{graphicx}
\usepackage[nosort]{cite}
\usepackage{epsfig,psfrag}

\makeatletter
\renewcommand\section{\@startsection {section}{1}{\z@}%
                                   {-3.5ex \@plus -1ex \@minus -.2ex}
                                   {2.3ex \@plus.2ex}%
                                   {\normalfont\large\bfseries}}
\renewcommand\subsection{\@startsection{subsection}{2}{\z@}%
                                     {-3.25ex\@plus -1ex \@minus -.2ex}%
                                     {1.5ex \@plus .2ex}%
                                     {\normalfont\bfseries}}
\makeatother

\theoremstyle{plain}

\theoremstyle{definition}

\let\non\nonumber

\let\d=\delta

\let\k=\kappa

\let\s=\sigma

\let\w=\wedge

\def\one{^{(1)}}

\newcommand{\bea}{\begin{eqnarray}}
\newcommand{\eea}{\end{eqnarray}}
\newcommand{\be}{\begin{equation}}
\newcommand{\ee}{\end{equation}}
\newcommand{\bma}{\begin{pmatrix}}
\newcommand{\ema}{\end{pmatrix}}

\newcommand{\hlf}{\frac{1}{2}}

\newcommand{\Z}{{\mathbb Z}}
\newcommand{\R}{{\mathbb R}}

\newcommand{\vol}{\operatorname{vol}}

\newcommand{\e}{\epsilon}

\newcommand{\m}{\mu}
\newcommand{\n}{\nu}
\newcommand{\p}{\partial}

\newcommand{\lp}{\left(}
\newcommand{\rp}{\right)}
\newcommand{\ls}{\left[}
\newcommand{\rs}{\right]}

\newcommand{\al}{\alpha}
\renewcommand{\d}{\delta}
\renewcommand{\k}{\kappa}
\newcommand{\om}{\omega}

\newcommand{\hph}[1]{{\hphantom{#1}}}

\newcommand{\C}[1]{$(\ref{#1})$}

\def\IZ{\relax\ifmmode\mathchoice
{\hbox{\cmss Z\kern-.4em Z}}{\hbox{\cmss Z\kern-.4em Z}}
{\lower.9pt\hbox{\cmsss Z\kern-.4em Z}} {\lower1.2pt\hbox{\cmsss
Z\kern-.4em Z}}\else{\cmss Z\kern-.4em Z}\fi}
\def\IR{\relax{\rm I\kern-.18em R}}

\def\one{{\hbox{ 1\kern-.8mm l}}}

\def\tr{{\rm tr\,}}

\newlength{\bredde}
\def\slash#1{\settowidth{\bredde}{$#1$}\ifmmode\,\raisebox{.15ex}{/}
\hspace*{-\bredde} #1\else$\,\raisebox{.15ex}{/}\hspace*{-\bredde}
#1$\fi}

\newsavebox{\zzzbar}
\sbox{\zzzbar}
  {\setlength{\unitlength}{0.9em}
  \begin{picture}(0.6,0.7)
  \thinlines
  \put(0,0){\line(1,0){0.6}}
  \put(0,0.75){\line(1,0){0.575}}
  \multiput(0,0)(0.0125,0.025){30}{\rule{0.3pt}{0.3pt}}
  \multiput(0.2,0)(0.0125,0.025){30}{\rule{0.3pt}{0.3pt}}
  \put(0,0.75){\line(0,-1){0.15}}
  \put(0.015,0.75){\line(0,-1){0.1}}
  \put(0.03,0.75){\line(0,-1){0.075}}
  \put(0.045,0.75){\line(0,-1){0.05}}
  \put(0.05,0.75){\line(0,-1){0.025}}
  \put(0.6,0){\line(0,1){0.15}}
  \put(0.585,0){\line(0,1){0.1}}
  \put(0.57,0){\line(0,1){0.075}}
  \put(0.555,0){\line(0,1){0.05}}
  \put(0.55,0){\line(0,1){0.025}}
  \end{picture}}

\newcommand{\ena}{\end{eqnarray}}
\newcommand{\beqa}{\begin{eqnarray}}
\newcommand{\eeqa}{\end{eqnarray}}

\def\bes #1\ees{\begin{split}#1\end{split}}

\newfont{\goth}{ygoth.tfm scaled 1200}                   

\def\e{\epsilon}

\def\k{\kappa}

\def\m{\mu}
\def\n{\nu}
\def\o{\omega}

\def\s{\sigma}

 \numberwithin{equation}{section}

\def\1{{(1)}}
\def\2{{(2)}}
\def\3{{(3)}}

\def\1{{\bf 1}}

\def\M{{\mathcal M}}

\def\1{{\bf 1}}
\def\3{{\bf 3}}
\def\7{{\bf 7}}
\def\2{{\bf 2}}
\def\8{{\bf 8}}

\begin{document}
\begin{titlepage}

\begin{center}

{September 11, 2013}
\hfill       \phantom{xxx}  EFI-13-20, DMUS-MP-13/18

\vskip 2 cm {\Large \bf New Examples of Flux Vacua}
\vskip 1.25 cm {\bf  Travis Maxfield$^{a}$\footnote{j.mcorist@surrey.ac.uk, \, maxfield@uchicago.edu, \, 
D.G.Robbins@uva.nl, \, sethi@uchicago.edu.},  Jock McOrist$^{b}$,  
Daniel Robbins$^{c}$ and Savdeep Sethi$^{a}$}\non\\
{\vskip 0.5cm  $^{a}${\it Enrico Fermi Institute, University of Chicago, Chicago, IL 60637, USA }\non\\
\vskip 0.2 cm
$^{b}${\it  Department of Mathematics, University of Surrey,
Guildford GU2 7XH, UK}\non\\
\vskip 0.2 cm
$^{c}${\it Institute for Theoretical Physics, University of Amsterdam,\\ Science Park 904, Postbus 94485, 1090 GL Amsterdam, The Netherlands}}

\end{center}
\vskip 2 cm

\begin{abstract}
\baselineskip=18pt
Type IIB toroidal orientifolds are among the earliest examples of flux vacua. 
By applying T-duality, we construct the first examples of massive IIA flux vacua with Minkowski space-times, along with new examples of type IIA flux vacua. 
The backgrounds are surprisingly simple with no four-form flux at all. They serve as illustrations of the ingredients needed to build type IIA and massive IIA solutions with scale separation. To check that these backgrounds are actually solutions, we formulate the complete set of type II supergravity equations of motion in a very useful form that treats the R-R fields democratically.  

\end{abstract}

\end{titlepage}


\section{Introduction}

The first and simplest examples of four-dimensional type IIB flux vacua that evade the classic supergravity no-go theorem of~\cite{Gibbons:1984kp}\ involve quotients of $T^6$~\cite{Dasgupta:1999ss}. The easiest fluxes to consider descend from classes on $T^6$ to the quotient space. The aim of this project is to examine the possible geometries that can be obtained from this starting point using T-duality on the ambient $T^6$.  This approach was originally employed to find torsional solutions of the type I/heterotic strings~\cite{Dasgupta:1999ss}, and subsequently used to construct N=2 type IIA flux vacua with dual Calabi-Yau descriptions~\cite{Schulz:2004tt}. However, a complete classification of the possible geometric backgrounds that can be found from this method has not been described. A scan of supersymmetric twisted tori vacua in the framework of generalized geometry can be found in~\cite{Grana:2006kf}. 

The reason for our interest in this exercise is that new examples of flux vacua for both type IIA and massive IIA can be found this way. For models with a non-zero Romans parameter $m$, this is particularly interesting because we have no prior examples of solutions to massive IIA with either Minkowski space-times, or $AdS_4$ space-times with the $AdS_4$ length scale significantly larger than the Kaluza-Klein scale. Indeed, massive IIA supergravity is one of the more mysterious corners of string theory~\cite{Romans:1985tz}. Classical vacua with $AdS_4$ space-times of Freund-Rubin type were described in the original work of Romans. Those solutions do not possess scale separation, and should not really be viewed as compactifications to four dimensions in the usual sense, but are really fully ten-dimensional backgrounds. More general classical $AdS_4$ solutions were described in subsequent work; see, for example~\cite{Behrndt:2004km, Tomasiello:2007eq, Petrini:2009ur, Lust:2009mb}. The possibility of classical IIA solutions with scale separation was discussed in~\cite{Tsimpis:2012tu}, but no examples are currently known. The status of various attempts to build type IIA and massive IIA flux vacua is summarized in~\cite{McOrist:2012yc}. 

Much like the torsional solutions of the heterotic/type I strings found in~\cite{Dasgupta:1999ss}, the new examples of type IIA and massive IIA flux vacua teach us about the topology of solutions that can satisfy the equations of motion and Bianchi identities, as well as the structure of necessary additional ingredients like orientifold planes and D-branes. The Romans mass parameter, $m$, of massive IIA is equivalent to a constant expectation value for the R-R $F_0$-flux in type IIA string theory. Because this flux does not dilute as we scale up the compactification volume, it is quite difficult to get a handle on solutions with scale separation from a direct analysis of the equations of motion expanded around a large volume solution. For example, a large volume Calabi-Yau space is not a good starting point for constructing a solution of massive IIA.  Indeed, it does not even appear to be a good starting point for constructing flux solutions of conventional type IIA string theory~\cite{McOrist:2012yc}. 

One might imagine that four-dimensional type IIA and M-theory flux compactifications are on better footing than massive IIA, since at least type IIA string theory has a perturbative quantum definition. Yet nothing could be further from the truth! Unlike type IIB string theory, the crucial ingredients needed to evade the classic supergravity no-go theorem have only recently been explored~\cite{McOrist:2012yc}, and are still very poorly understood. The central ingredient appears to be the ability of orientifold planes and D-branes to generate brane charge in directions normal to their world volume. A systematic understanding of the couplings supported on D-branes and orientifold planes, and the charges they can produce, is sorely needed information. Recent work on higher derivative interactions, including those supported on branes, can be found in~\cite{Garousi:2009dj, Garousi:2010bm, Garousi:2010ki, Becker:2010ij, Garousi:2010rn, Becker:2011bw, Becker:2011ar, Garousi:2011fc, Garousi:2011ut, Basu:2011he, McOrist:2012yc, Garousi:2012yr, Garousi:2012jp, Hatefi:2012rx, Hatefi:2012ve, Hatefi:2012wj, Hatefi:2012zh, Hatefi:2013mwa, Liu:2013dna, Godazgar:2013bja, Garousi:2013lja, Basu:2013goa, Basu:2013oka}. 

Let us summarize our main observations as we outline this paper. In section~\ref{newgeom}, we begin by ignoring the quotient action entirely. Instead, we work with a torus geometry for which we classify all possible fluxes compatible with the equations of motion. This includes fluxes that both preserve and break supersymmetry. Inclusion of SUSY breaking fluxes turns out to be crucial if one wishes to generate a non-zero Romans mass from T-duality. We then describe the constraints that need to be imposed on the choice of fluxes so that T-duality gives back an honest geometry, rather than a quantum geometry requiring T-duality to patch open sets. The basic condition is that we T-dualize only once along any $3$-cycle supporting an $H_3$-flux~\cite{Shelton:2005cf}. It would be interesting to relax this condition and consider more general T-folds, but we will restrict to geometry in this work. 

We then classify the distinct ways of performing $3$ T-dualities on $T^6$. The resulting vacua are remarkably nice with the simplest massive IIA example still a topological $6$-torus with a metric of schematic form:
\be
ds_{IIA}^2 = e^{-w} g_{\mu\nu} dx^\mu dx^\nu + e^{w} ds^2_{T^3}  + e^{-w} ds^2_{{\widetilde T}^3}. 
\ee
Here $w$ is the original type IIB warp factor. Up to an additive constant, the type IIA dilaton is varying and also determined by the warp factor $\phi_A = -{3\over 2}w$. There is a precisely correlated pair of fluxes $(F_0, H_3)$, and an additional $F_2$-flux, but no $F_4$-flux. Indeed, none of the examples we find in either type IIA or massive IIA has $4$-form flux.\footnote{Orientifold examples of type IIA and M-theory vacua with $4$-form flux can be found in~\cite{McOrist:2012yc}. These examples are constructed using $1$ T-duality rather than the $3$ T-dualities studied here.} Related backgrounds have been studied at the level of supergravity or effective field theory, without explicit solutions, in past work~\cite{Lust:2008zd, Camara:2007cz}. However, those analyses do not include the ingredients needed to evade the usual supergravity no-go theorem for flux compactifications in type IIA string theory. The inclusion of those necessary ingredients, which we can determine from T-duality, permits solutions to the dilaton and warp factor equations on a compact space, while also giving tadpole constraints on the choice of fluxes.   

Supersymmetry is always broken in the massive IIA case, but not necessarily in the type IIA case. In general, the massive IIA vacua are not topologically $6$-tori, while the pure type IIA flux vacua are never $6$-tori, as we discuss in section~\ref{cohomology}, but it is very nice to have one simple case where this is true. 

In section~\ref{oplanes}, we consider the effects of two different quotient actions and track the action of T-duality on the original array of type IIB planes and branes. In the resulting T-dual type IIA theories, we only find $O6$-planes and D6-branes. However, we do find that generically the $O6$-planes and D6-branes contribute to a D6-brane charge tadpole in an unexpected way: they induce D6-brane charge in a direction normal to their own charge. This is true both in conventional type IIA and massive IIA. This appears to be a simple example of the phenomenon described in~\cite{McOrist:2012yc}; 
we hope to explore how this works in more detail elsewhere. 

It is also important to stress that we are not using any strange ingredients like smeared orientifolds that have been studied in past work like~\cite{Andriot:2010ju, Blaback:2010sj, Blaback:2012mu, Caviezel:2008ik, Petrini:2013ika}.  All the ingredients appearing in our backgrounds are robust in string theory. We should add that this does not mean  flat non-compact $O6$-planes exist in massive IIA. What we find in massive IIA are wrapped $O6$-planes in conjunction with specific fluxes and possibly D6-branes. This specific combination of ingredients do appear to exist in massive IIA.\footnote{Some potentially smooth geometries in massive IIA that  resemble $O6$-planes have been studied in~\cite{Saracco:2012wc}. See comments in the last section of~\cite{McOrist:2012yc}.}

Section~\ref{demfluxes}\ is devoted to a formulation of the type II supergravity equations of motion in a way that treats the R-R fields democratically.  This is a very convenient form for checking that our backgrounds actually solve the equations of motion. This check can be performed at the level of supergravity, without worrying about the contribution of any beyond supergravity ingredients like orientifolds, branes or higher derivative interactions for two reasons: first, one can consider non-compact solutions for which such ingredients are not needed, and these backgrounds solve the SUGRA equations of motion on the nose. Second, for compact solutions such ingredients contribute in a known way to the type IIB equations of motion. The type IIA image of those contributions can therefore be determined by T-duality. 

In section~\ref{simple}, we take a closer look at these backgrounds. Specifically, we examine the D6-brane tadpole equation and we check that the equations of motion are satisfied. It is interesting to see how the IIA and massive IIA equations are solved since it gives us a starting point for formulating a more general IIA ansatz. There are many interesting questions to ponder; for example,  do supersymmetric flux vacua exist in massive IIA with Minkowski space-times? Can type IIA $AdS_4$ solutions be found with scale separation?  Can a more general IIA metric ansatz be formulated that allows all fluxes to appear?  How do $O6$-planes and D6-branes generate D6-brane charge in unusual directions?  Can a four-dimensional effective field theory description be found for type IIA and M-theory flux backgrounds? We hope to explore some of these questions in future work. 





\section{New Geometries from T-duality}\label{newgeom}

\subsection{Fluxes on tori}

One of the nicest examples of a type IIB flux solution is based on the geometry $T^2\times K3$~\cite{Dasgupta:1999ss}. It is important that the space not be flat in order to solve the D3-brane tadpole condition. Different choices of $G_3$-flux preserve either $N=0, 1$ or $2$ space-time supersymmetries. For our goal of studying the geometries that emerge from duality, we can replace $K3$ by a quotient of $T^4$ and track the quotient action later. We can also consider more general quotient actions on $T^6$. Indeed, we might as well make the geometry as simple as possible and consider $T^2\times T^2\times T^2$ as our starting point. Let us take complex coordinates $(x,y,z)$ for this space with $x=x_1+ix_2$ and $x_i \sim x_i+1$ etc. Choose the square complex structure for each $T^2$ factor.  The starting type IIB metric takes the form:
\be\label{iibmetric}
e^{-{\phi_B \over 2}}ds_{IIB}^2 = e^{-w} g_{\mu\nu} dx^\mu dx^\nu + e^{w} \left( dx d\bar x + dyd\bar y + dz d\bar z \right). 
\ee
The warp factor, $w$, is only a function of the internal coordinates $(x,y,z)$. The type IIB string coupling, $\tau_B = C_0 + i e^{-\phi_B}$, is independent of $(x,y,z)$. We will set $C_0=0$ for simplicity.  

The type IIB background also includes a self-dual $F_5$-flux. The mostly space-time components of $F_5$ are given by
\be\label{F5spacetime}
F_5^{space} = -d\left(e^{-2w}\right)\wedge \vol_4,
\ee
where $\vol_4$ is the volume form of the unwarped space-time metric. See Appendix~\ref{signconventions}\ for sign conventions, and~\C{volumefour}\ for a definition of $\vol_4$.
In order to correctly T-dualize the R-R fluxes, it is going to be simpler to work with the R-R field strengths rather than the R-R potentials. The Bianchi identity for the self-dual ${F}_5$ leads to a Laplace type equation for the warp factor: 
\be\label{warp}
d\ast {F}_5 = d{F}_5 = - H_3\wedge F_3 + X_6 + \sum_i \delta^2(x-x_i)\delta^2(y-y_i)\delta^2(z-z_i). 
\ee
The gravitational source, $X_6$, arises from higher derivative interactions supported on $D$-branes and orientifold planes; for example, one source is the $\int C_4\wedge \tr( R\wedge R)$ coupling supported on $(p,q)$ $7$-branes.  Without this higher derivative contribution, no compact solution is possible. The last term in~\C{warp}\ is the contribution from any space-filling D3-branes present in the background. 

From the perspective of M-theory, we are considering $T^2\times T^2\times T^2\times T^2$ with complex coordinates $(u,x,y,z)$, and a square complex structure for the $(x,y,z)$ tori. The $u$-torus is distinguished with a complex structure determining the type IIB string coupling, $\tau_u=\tau_B=ie^{-\phi_B}$ and $u=u_1+ie^{-\phi_B}u_2$. The type IIB limit requires taking the area of the $u$-torus to zero. We could generalize this starting point by allowing a more general complex structure, but it will obscure the main issues we want to explore. 

Now we want to parametrize the most general choice of $G_3$-flux that can solve the equations of motion. A nice approach is to start in M-theory on $T^2\times T^2\times T^2\times T^2$ and consider all $G_4$-flux that is self-dual: $G_4=\ast G_4$~\cite{Becker:1996gj}. The cohomology $H^4(T^8, \Z)$ contains $70$ elements. Of these classes, $35$ are self-dual and can be decomposed by Hodge type: they are combinations of the $(4,0)$ and $(0,4)$ elements, $12$ non-primitive combinations of $(3,1)$ and $(1,3)$, $20$ primitive $(2,2)$ classes together with the square of the K\"ahler form, which is $(2,2)$ and non-primitive.  

Of these $35$ fluxes, the only choices that are compatible with the lift to a Lorentz invariant type IIB background are those with one leg along the $u$-torus~\cite{Dasgupta:1999ss}. There are $20$ such choices. We can provide a basis for these forms, 
\bea\label{basicchoices}
& dudxdydz ,  \,\, d\bar u d\bar x dydz, \,\, d\bar u dx d\bar y dz,  \,\,  d\bar u dxdy d\bar z,  & \cr
& d\bar u dx \left( dyd\bar y -  dzd\bar z \right), \,\, d\bar u dy \left( dzd\bar z - dx d\bar x \right), \,\,  d\bar u dz \left( dyd\bar y -  dxd\bar x \right), & \cr
& du dx \left(dyd\bar y + dz d\bar z\right), \, \  du \left(dx d\bar x + dyd\bar y \right) dz, \, \  du \left(dx d\bar x + dz d\bar z\right) dy, &
\eea
together with their complex conjugates. Not all of these choices will be compatible with the desired quotient action, or the tadpole condition, but we will worry about those constraints later. 

The most general $G_4$ we will consider is a real linear combination of the forms~\C{basicchoices}\ and their complex conjugates together with some integrality constraint. Denote the ordered list of $4$-forms appearing in~\C{basicchoices}\ by  $\left\{ \o_1, \ldots, \o_{10} \right\}$. We can express $G_4$ as follows, 
\be\label{fluxparameters}
{G_4 \over 2\pi} = \sum_{i=1}^{10} a_i \left(\o_i + \bar\o_i\right)+ i b_i \left(\o_i - \bar\o_i\right),
\ee
where the choice of $(a_i, b_i)$ determines a choice of flux. 
From $G_4$, we can read off the corresponding type IIB $H_3$ and $F_3$ fluxes using
\be
{G_4 \over 2\pi} = H_3 du_1 + F_3 du_2. 
\ee
Space-time supersymmetry is broken by either turning on fluxes with a $(4,0)$ component, or a non-primitive $(3,1)$ component. These fluxes correspond to the $(a_1, b_1)$ coefficient, and to the $(a_8, a_9, a_{10})$ and $(b_8, b_9, b_{10})$ coefficients. These two distinct ways of breaking supersymmetry correspond to $F$-term or $D$-term breaking, respectively, in four dimensions. 

If we want to restrict to $F$-term breaking in four dimensions, we need to impose a primitivity condition with respect to the canonical K\"ahler form: 
\be
J = {i\over 2}\left( dud\bar u +dxd\bar x+dyd\bar y+dzd\bar z \right). 
\ee 
This constraint is not needed to solve the flux equations of motion.  Imposing $J\wedge \omega=0$ for a self-dual $\omega\in H^4(T^8, \Z)$ with a leg along the $u$-torus leaves a total of $14$ possibilities. In the expansion~\C{fluxparameters}, this constraint sets:
\be\label{ftermbreaking}
a_8=a_9=a_{10}=0, \qquad b_8=b_9=b_{10}=0.
\ee
Equivalently, the last line of~\C{basicchoices}\ is set to zero. 

\subsection{T-duality}
\label{tdualcomments}
To determine the metric that results from applying T-duality, we only need to worry about the $H_3$-flux. We can ignore the R-R flux and the dilaton.  The main constraint is that we only want to generate geometric backgrounds that do not require patching conditions involving T-duality. This means we can only T-dualize once along any $3$-cycle supporting $H_3$-flux. Without specifying a quotient action, the directions $(x,y,z)$ are completely symmetric so we are free to T-dualize along $x_1$ as a first step. Let us examine the additional T-dualities that we might consider in the search for a combination that can generate an $F_0$-flux. We will need at least $3$ T-dualities if we have any hope of converting an $F_3$-flux into an $F_0$-flux. More than three T-dualities leads to a non-geometric background so we will consider three T-dualities. 

Some general comments about T-dualizing R-R fluxes are appropriate at this point: the R-R field strength is a self-dual object. Our starting type IIB configuration has $(F_3, F_5, F_7)$ fluxes, where $F_7$ is Hodge dual\footnote{Precise relations between R-R fluxes and their duals can be found in Appendix~\ref{signconventions}. } to $F_3$.  The $F_5$ flux is self-dual with two components: the first is space-time filling with one internal leg given in~\C{F5spacetime}, while the second component has five internal legs. The second component is the ten-dimensional Hodge dual of the first component. We can denote these two components $(F_5^{space}, F_5^{int})$. 

T-duality preserves self-duality of the R-R field strength. After three T-dualities, the resulting IIA R-R field strength will contain contributions from the $F_3$ and $F_7$ fluxes which are Hodge dual, and contributions from $(F_5^{space}, F_5^{int})$, which are also Hodge dual. We only need to specify one half of the components of the resulting R-R field strength. We will choose to specify $(F_0, F_2, F_4)$. To find these fluxes, we can T-dualize the original type IIB $F_3$ and $F_5^{space}$ fluxes; the $F_5^{space}$ flux will generate fluxes of the generic form $(F_6, F_8)$ which can be Hodge dualized to $(F_0, F_2, F_4)$. The $F_3$-flux can generate $(F_0, F_2, F_4, F_6)$, which can again be dualized to  $(F_0, F_2, F_4)$. It is much simpler to consider field strengths rather than potentials in implementing T-duality. 


\subsubsection{Dualizing $(x_1, x_2)$}\label{firstcase}

Let us start by imposing the constraint~\C{ftermbreaking}\ that all supersymmetry breaking is by an $F$-term. 
If we choose to perform a second T-duality along $x_2$, we must supplement~\C{ftermbreaking}\ by setting to zero: 
\be a_6=b_6=0, \qquad a_7=b_7=0. \ee 
This leaves $12$ potential choices of $H_3$-flux. However, T-duality along $(x_1, x_2)$ turns all of these choices of $H_3$-flux into metric. It also converts the R-R $F_3$-flux into $F_3$-flux. 

This is the case first considered in~\cite{Dasgupta:1999ss}\ which leads to a torsional solution of the type I string. The $x$-torus is fibered over the $(y,z)$ directions with a fibration determined by the $H_3$-flux. We cannot generate an $F_0$-flux this way so we must either allow $D$-term supersymmetry breaking, or consider a second T-duality along the $y$ or $z$ directions.

\subsubsection{Dualizing $(x_1,x_2,y_1)$}

Let us first relax the constraint~\C{ftermbreaking}\ and allow $D$-term breaking fluxes. Symmetry singles out no special direction for the third T-duality so we may as well choose $y_1$. Imposing the condition that we end up with an honest geometry produces $10$ linear constraints on the flux parameters $(a_i, b_i)$ given below: 
\bea
& a_1 + a_3 = a_2 + a_4 = a_5 + a_8 = a_7 - a_9 = a_6 - a_{10} = 0,  &\cr
& b_1 + b_3 = b_2 - b_4 = b_5 + b_8 = b_7 - b_9 = b_6 - b_{10} = 0.&
\eea
The flux is then determined by $10$ parameters, which we can take to be $(a_1, a_2, a_5, a_6, a_7)$ and $(b_1, b_2, b_5, b_6, b_7)$. The $H_3$-flux is given explicitly by,
\bea\label{hfluxone}
H_3 &=& 4a_1( dx_1dy_2dz_2 + dx_2dy_2dz_1) +4a_2(dx_1dy_2dz_2 - dx_2dy_2dz_1) +8a_5dx_2dz_1dz_2 \cr && -8a_6dy_2dz_1dz_2 - 8a_7dy_1dy_2dz_2 +4b_1(dx_1dy_2dz_1 - dx_2dy_2dz_2) \cr && +4b_2( dx_1dy_2dz_1 + dx_2dy_2dz_2) + 8b_5dx_1dz_1dz_2 - 8b_6dy_1dz_1dz_2 \cr && -8b_7dy_1dy_2dz_1.
\eea
This now has components that survive the $3$ T-dualities as $H_3$-flux. The $F_3$-flux is given by, 
\bea
e^{\phi_B}F_3 &=& 4a_1( dx_1dy_1dz_2 + dx_2dy_1dz_1) +4a_2(dx_2dy_1dz_1 - dx_1dy_1dz_2) +8a_5dx_1dy_1dy_2 \cr && -8a_6dx_1dx_2dy_1 - 8a_7dx_1dx_2dz_1 +4b_1(dx_1dy_1dz_1 - dx_2dy_1dz_2) \cr && -4b_2( dx_1dy_1dz_1 + dx_2dy_1dz_2) - 8b_5dx_2dy_1dy_2 + 8b_6dx_1dx_2dy_2 \cr && +8b_7dx_1dx_2dz_2.
\eea
The coefficient $a_6$ turns on a component of $F_3$ proportional to $dx_1dx_2dy_1$, which can dualize to an $F_0$-flux. Notice that this is correlated with a component of $H_3$-flux in~\C{hfluxone}\ which survives T-duality. However, $a_6 = a_{10}$ turns on a  non-primitive flux, as we expected from section~\ref{firstcase}, breaking supersymmetry via a $D$-term.

\subsubsection{Dualizing $(x_1,y_1, z_1)$} \label{x1y1z1}

Without some choice of quotient action singling out particular directions, the only other distinct possibility is dualizing $(x_1, y_1, z_1)$. This is the case we will focus on for the remainder of this paper. 

Imposing the constraint that we end up with a geometry, we find that the $H_3$-flux is determined by $10$ parameters. Let us first specify the M-theory $G_4$-flux,
\bea
&b_1= - b_2= - b_3 =- b_4, \quad b_5= b_6=b_7=b_8=b_9=b_{10}=0, &\cr 
& a_1+a_2+a_3+a_4=0.&
\eea
We can choose the $10$ flux parameters to be $b_1$ and $(a_2, \ldots, a_{10})$. Writing out the $H_3$-flux explicitly gives:
\bea \label{hflux}
H_3 &=& - 4a_2 (dx_2dy_1dz_2 +  dx_2dy_2dz_1) - 4a_3(dx_1dy_2dz_2  +dx_2dy_2dz_1) - 4 a_4( dx_1dy_2dz_2  \cr && +dx_2dy_1dz_2 )   + 4a_5(dx_2dz_1dz_2  -  dx_2dy_1dy_2) + 4a_6 (dx_1dx_2dy_2 - dy_2dz_1dz_2)     \cr && + 4 a_7(dx_1dx_2dz_2- dy_1dy_2dz_2) -4a_8(dx_2dz_1dz_2 + dx_2dy_1dy_2) \cr &&-4a_9(dx_1dx_2dz_2 + dy_1dy_2dz_2)  -4a_{10}( dx_1dx_2dy_2 + dy_2dz_1dz_2) \cr && - 8b_1 dx_2dy_2dz_2.  
\eea
In a similar way, we can write out the starting $F_3$-flux:
\bea \label{fflux}
e^{\phi_B}F_3 &=& - 4a_2 (dx_1dy_1dz_2 +  dx_1dy_2dz_1) - 4a_3(dx_1dy_1dz_2  +dx_2dy_1dz_1) - 4 a_4( dx_1dy_2dz_1  \cr && +dx_2dy_1dz_1 )   + 4a_5(dx_1dy_1dy_2  -  dx_1dz_1dz_2) + 4a_6 (dy_1dz_1dz_2 - dx_1dx_2dy_1)     \cr && + 4 a_7(dy_1dy_2dz_1- dx_1dx_2dz_1)  -4a_8(dx_1dy_1dy_2 + dx_1dz_1dz_2) \cr && -4 a_9(dx_1dx_2dz_1 + dy_1dy_2dz_1) - 4a_{10}(dx_1dx_2dy_1 + dy_1dz_1dz_2) \cr &&+ 8b_1 dx_1dy_1dz_1.  
\eea
Notice that only  $b_1$ gives a term in $F_3$ of the form $dx_1dy_1dz_1$, which can become $F_0$-flux. However, $b_1$ corresponds to turning on a $(4,0)$ flux in M-theory or a $(3,0)$ flux in type IIB. Such a flux breaks space-time supersymmetry spontaneously via an $F$-term. 

\subsubsection{The dual theory for the simplest example}\label{simplest}

Before looking at the details of the metric that emerges from dualizing $(x_1, y_1, z_1)$ in the general case, let us note that all of the $H_3$-flux, other than the coefficient of $b_1$, turns into metric. The only surviving $H_3$-flux is again precisely the component correlated with the appearance of $F_0$-flux. 

The simplest example is therefore to set $a_2=\ldots = a_{10}=0$ and consider only $b_1 \neq 0$, with only $F$-term supersymmetry breaking.  In this case, we find a type IIA string frame metric of the form:
\bea
ds_{IIA}^2 &=& e^{-w + {\phi_B\over 2}} g_{\mu\nu} dx^\mu dx^\nu + e^{w+ {\phi_B\over 2}} \left\{ (dx_2)^2  +  (dy_2)^2 +  (dz_2)^2 \right\}  \cr && + e^{-w- {\phi_B\over 2}} \left\{ (dx_1)^2  +  (dy_1)^2 +  (dz_1)^2 \right\}. 
\eea
The $H_3$-flux and type IIA dilaton are given by:
\be\label{simpleNS}
H_3 = - 8b_1 dx_2dy_2dz_2, \qquad \phi_A = {\phi_B \over 4} - {3\over 2} w.
\ee 
Notice that $\phi_A$ is varying because the warp factor depends on the internal coordinates. This space is topologically $T^6$. It is metrically $T^3 \times T^3$. 

Now we can turn to the R-R fluxes. We will follow the strategy described in section~\ref{tdualcomments}. The first contribution comes from dualizing $F_3$ of~\C{fflux}\ using the rules in Appendix~\ref{tduality}. The $B_2$ potential associated to the $H_3$-flux has no legs in the $(x_1,y_1,z_1)$ directions so T-duality acts straightforwardly giving:
\be
F_0 = -8b_1e^{-\phi_B}. 
\ee
There is one other contribution which comes from the mostly space-time filling $F_5^{space}$ field strength of~\C{F5spacetime}. T-duality is again straightforward giving,
\be
F_8 =- d\left(e^{-2w} \right) \wedge \vol_4 \wedge dx_1 \wedge dy_1 \wedge dz_1.
\ee
This $F_8$ field strength dualizes to an $F_2$ field strength supported on the internal space:
\be
{F}_2=-\ast F_8=\ast_3d_3\lp e^{2w}\rp.
\ee
The Hodge star, $\ast_3$, is with respect to the unwarped metric, while $d_3$ acts only on the three-torus given by $(x_2,y_2,z_2)$.   It is rather surprising that there is such a simple solution to the Romans theory with only $(F_0, F_2, H_3)$ fluxes.  We will return to this example later in section~\ref{simple}\ with the aim of examining how the equations of motion are being solved. 

\subsubsection{The dual theory for the general case}\label{generalcase}

The general case involves ``geometric flux'' or non-trivial circle bundles. For a review of twisted tori vacua, including very closely related metrics to those we find below, see~\cite{Andriot:2010ju}. We first need to choose a trivialization of $H_3$. There are many gauge-equivalent choices, and one such choice appears below:
\bea \label{bflux}
B_2 &=& - 4a_2 x_2 (dy_1dz_2 +  dy_2dz_1) + 4a_3 y_2 (dx_1dz_2  +dx_2dz_1) - 4 a_4 z_2 ( dx_1dy_2  \cr && +dx_2dy_1 )   + 4a_5(z_2 dx_2dz_1  -  y_2 dx_2dy_1) + 4a_6 (y_2 dx_1dx_2 -z_2  dy_2dz_1)     \cr && + 4 a_7 z_2 (dx_1dx_2- dy_1dy_2) -4a_8x_2(dz_1dz_2 + dy_1dy_2) \cr &&-4a_9z_2(dx_1dx_2 + dy_1dy_2)  -4a_{10}y_2( dx_1dx_2 + dz_1dz_2) \cr && - 8b_1 x_2dy_2dz_2, \\
&=&  A_x dx_1 + A_y dy_1 + A_z dz_1  - 8b_1 x_2dy_2dz_2.
\eea
The potentials $(A_x, A_y, A_z)$ determine the twisting of the $(x_1, y_1, z_1)$ circles over the base $T^3$ with coordinates $(x_2, y_2, z_2)$.  The dual metric follows from the rules of Appendix~\ref{tduality}, 
\bea\label{twistedmetric}
ds_{IIA}^2 &=& e^{-w + {\phi_B\over 2}} g_{\mu\nu} dx^\mu dx^\nu + e^{w+ {\phi_B\over 2}} \left\{ (dx_2)^2  +  (dy_2)^2 +  (dz_2)^2 \right\}  \cr && + e^{-w- {\phi_B\over 2}} \left\{ (dx_1+A_x)^2  +  (dy_1+A_y)^2 +  (dz_1+A_z)^2 \right\}. 
\eea
The $H_3$-flux and dilaton are still given by~\C{simpleNS}. This space is no longer topologically a $T^6$  as long as at least one of the potentials $(A_x, A_y, A_z)$ is non-zero. 

The R-R fluxes dualize to $F_0=-8b_1e^{-\phi_B}$, as we saw in the simplest case described in section~\ref{simplest}. In addition we now find R-R $2$-form flux from dualizing~\C{fflux}, 
\bea \label{generalRR}
e^{\phi_B}F_2^{flux} &&= 4\ls (a_3+a_4)dx_2+(a_7-a_9)dy_2+(-a_6+a_{10})dz_2\rs\wedge\lp dx_1+A_x\rp\non\\
&& +4\ls (a_7+a_9)dx_2+(a_2+a_4)dy_2-(a_5+a_8)dz_2\rs\wedge\lp dy_1+A_y\rp\\
&& +4\ls (-a_6-a_{10})dx_2+(-a_5+a_8)dy_2+(a_2+a_3)dz_2\rs\wedge\lp dz_1+A_z\rp. \non
\eea
Surprisingly, there is no $F_4$ generated by dualizing~\C{fflux}. 
The mostly space-time $F_5^{space}$ of~\C{F5spacetime} becomes an $8$-form field strength,
\be
F_8 = F_5^{space} \wedge\lp dx_1+A_x\rp\wedge\lp dy_1+A_y\rp\wedge\lp dz_1+A_z\rp.
\ee
%
Taking the Hodge dual gives, 
\be\label{partialflux}
{\widetilde F}_2=-\ast F_8=\ast_3d_3\lp e^{2w}\rp,
\ee
where $\ast_3$ is again the Hodge star with respect to the unwarped metric, while $d_3$ again acts only on the three-torus given by $(x_2,y_2,z_2)$.  
%

Let us summarize the final data for this general case. The metric, which is of twisted torus type, appears in~\C{twistedmetric}, while the fluxes and type IIA dilaton appear below:
\bea\label{generalfluxes}
H_3 = - 8b_1 dx_2dy_2dz_2, \qquad F_0=-8b_1e^{-\phi_B}, \\ F_2=F_2^{flux} +\ast_3d_3\lp e^{2w}\rp,  \qquad \phi_A = {\phi_B \over 4} - {3\over 2} w.
\eea
We can list our major observations: 
\begin{itemize}
\item The lack of any $F_4$-flux. 
\item The Romans parameter is completely correlated with $H_3$. 
\item Supersymmetry is spontaneously broken. 
\end{itemize}

\subsection{Cohomology}\label{cohomology}

While the simplest example of section~\ref{simplest}\ is still topologically $T^6$, this is not true for the general case with metric~\C{twistedmetric}. This is a well known consequence of ``geometric flux'' or the generation of non-trivial circle bundles from dualizing $H_3$,  but worth iterating here~\cite{Dasgupta:1999ss, Kachru:2002he, Becker:2009df}. The Betti numbers for $T^6$, satisfying $b_n=b_{6-n}$,  are given below:
\be
b_0=1, \quad b_1=6, \quad b_2=15, \quad b_3=20. 
\ee
When a non-trivial gauge potential like $A_x$ appears in~\C{twistedmetric},  the form $dx_1$ is no longer globally defined if the metric is to be well-defined. Rather the periodicity of $x_1$ is modified so that the combination,
\be
dx_1 + A_x,
\ee
is globally defined, but not closed. Therefore $b_1$ is reduced by $1$. This is sufficient to see topology change. For similar reasons, $b_2$ and $b_3$ change. 
For example, the field strength $dA_x$ is an exact $2$-form supported along the $(x_2, y_2, z_2)$ directions. This reduces $b_2$ by $1$. Similarly,
$$
(dx_1 + A_x)\wedge dy_1, \qquad (dx_1 + A_x)\wedge dz_1,
$$ 
are no longer closed, reducing $b_2$ further by $2$. The field strength $dA_x$ defines a linear map, 
\be
dA_x: \Omega^1(S^1_{x_2}\times S^1_{y_2}\times S^1_{z_2}, \R) \rightarrow \Omega^3(S^1_{x_2}\times S^1_{y_2}\times S^1_{z_2}, \R),
\ee
via the wedge product. The kernel of this map is $2$-dimensional. Let ${\alpha}$ denote a class in the one-dimensional complement of the kernel then
$$ 
(dx_1 + A_x)\wedge \alpha
$$
is also no longer closed. In total, $b_2$ decreases from $15$ to $11$ for this example. A similar analysis can be performed for $b_3$, and extended to the case of  any three specific gauge potentials  $(A_x, A_y, A_z)$. 
The central observation for us is that the space is only topologically $T^6$ for the special case of the simplest example with a non-vanishing Romans parameter described in section~\ref{simplest}. 


\section{Quotients, Branes and Orientifold-planes}\label{oplanes}

To this point, we have worked with the ambient $T^6$ without worrying about the choice of quotient action, satisfying tadpole constraints, or tracking branes and orientifold planes.  Let us examine each of these issues.

\subsection{Quotient actions}

All the solutions with a non-vanishing Romans parameter are non-supersymmetric. We can therefore consider various possible quotient actions that give orbifolds of $T^8$ in M-theory, or orientifolds of $T^6$ in type IIB, which  preserve different amounts of supersymmetry. Let ${\M}=T^8/G$ denote the M-theory orbifold obtained by quotienting $T^8$ by the finite group $G$. We would like the geometry to preserve some supersymmetry so the breaking by the flux looks spontaneous from the perspective of three or four-dimensional effective field theory.   When formulated in the language of M-theory, the $M2$-brane or $D3$-brane charge tadpole condition reads~\cite{Becker:1996gj, Sethi:1996es}:
\be \label{tadpole}
\# {\rm branes} + {1\over 2} \int { {G_4\over 2\pi} \wedge {G_4\over 2\pi} } = {\chi(\M) \over 24}. 
\ee
For models with type IIB limits, which are the cases of interest to us, the class ${G_4\over 2\pi} \in H^{4}(\M, \Z)$~\cite{Witten:1996md}. Computing the flux contribution to~\C{tadpole}\ from the our general expression~\C{fluxparameters}\ for the models involving dualization along $(x_1, y_1, z_1)$ discussed in section~\ref{x1y1z1}\ gives, 
\be
{1\over 2} \int { {G_4\over 2\pi} \wedge {G_4\over 2\pi} } = 32 e^{-\phi_B} \left[ 2 (b_1)^2 + a_2a_3 + a_2 a_4 +a_3 a_4 +\sum_{i=2}^{10} (a_i)^2\right] \times {1\over {\rm ord(G)}}.
\ee
Let us now examine two specific examples found in~\cite{Dasgupta:1999ss}. 

\subsubsection{$T^8/\Z_2$}
This is a $\Z_2$ quotient denoted ${\cal I}$ that acts as follows:
\be
{\cal I}: (u,x,y,z) \rightarrow (-u,-x,-y,-z). 
\ee
This is a space with terminal singularities. Nevertheless, the charge tadpole can be computed from the string orbifold definition of the Euler characteristic giving ${\chi(\M) \over 24} = 16$. All the fluxes appearing in~\C{fluxparameters}\ are invariant under this quotient action. This leaves many ways to satisfy the  tadpole condition. 

Let us look at the simplest example with $b_1\neq 0$ and $a_2=\ldots=a_{10}=0$. For simplicity, set $e^{\phi_B}=1$. The flux is given by,
\be\label{b1only}
{G_4\over 2\pi} = 8 b_1 \left( du_1dx_2dy_2dz_2 - du_2dx_1dy_1dz_1\right). 
\ee
The volume of $4$-cycles on the quotient space is reduced by a factor of $2$. The flux quantization condition becomes $4 b_1 \in \Z$ while the tadpole condition becomes:
\be
2 (4b_1)^2 +\# {\rm branes}= 16.
\ee
For this simplest case, we can choose $4b_1=2$ then tadpole cancelation requires an additional $8$ $M2$-branes.
 
If we want to cancel the tadpole with only flux, we need to turn on some $a$ parameters in~\C{fluxparameters}. For example, turning on $a_2$ and $a_8$ gives a $4$-form flux:
\bea\label{a2a8}
{G_4\over 2\pi} &=& 8 b_1 \left( du_1dx_2dy_2dz_2 - du_2dx_1dy_1dz_1\right)\cr && +4a_2 \left( du_1dx_2dy_1dz_2 + du_1dx_2dy_2dz_1 + du_2dx_1dy_1dz_2+du_2dx_1dy_2dz_1\right) \cr && +4a_8 \left( du_1dx_2dy_1dy_2 + du_1dx_2dz_1dz_2 + du_2dx_1dy_1dy_2+du_2dx_1dz_1dz_2\right). 
\eea
 For these coefficients $2a_i\in \Z$. The tadpole condition becomes, 
 \be
 2 (4b_1)^2 + 4(2a_2)^2 + 4(2a_8)^2 = 16.
 \ee
 Choosing $4b_1=2, \,\, 2a_2=2a_8=1$ solves the tadpole condition completely by flux without the need for additional $M2$-branes. However, before quotienting, the dual theory will now be of twisted torus type rather than topologically a $T^6$ space. 

This example lifts to type IIB on $T^6/(\Omega (-1)^{F_L}\Z_2)$ where the $\Z_2$ involution inverts $(x,y,z)$ producing $64$ $O3$-planes. This orientifold was also studied in~\cite{Kachru:2002he}. After T-dualizing along $(x_1,y_1,z_1)$, each $O3$-plane becomes an $O6$-plane generated by the  orientifold action $\Omega (-1)^{F_L} {\cal I'}$, where ${\cal I'}$ is the following $\Z_2$ involution:
\be
{\cal I'}: (x_2, y_2, z_2) \rightarrow (-x_2, -y_2, -z_2). 
\ee
This is a nice clean example but it suffers from the presence of naked orientifold planes in string theory, or equivalently, terminal singularities in M-theory. Neither description is really well controlled in supergravity; see~\cite{McOrist:2012yc}\ for a discussion of this issue. 

\subsubsection{$K3\times K3$}

A nicer example that does not involve naked orientifold planes is obtained by considering $K3\times K3$. Generically, this is a smooth compactification of M-theory. We can replace the first $K3$ by $T^4/\Z_2$, where the $\Z_2$ action inverts $(u,x)$, in order to obtain the perturbative type IIB orientifold, 
\be
{T^2\over \Omega (-1)^{F_L} \Z_2 }\times K3, 
 \ee
where the $\Z_2$ action inverts the $x$-torus. This generates $4$ fixed points, each supporting a single $O7$-plane together with $4$ D7-branes. There is no net D7-brane charge. The cancelation between the $O7$-plane and the D7-branes is pointwise. To perform our duality transformation, we would like to smoothly replace the $K3$ surface by an orbifold $T^4/\Z_2$. In M-theory, this means we consider the quotient of $T^8$ by the group with generators:
\be\label{mquotient}
{\widetilde g}_1:  (u,x) \rightarrow (-u,-x), \qquad g_2: (y,z) \rightarrow (-y,-z).
\ee
From the perspective of type IIB, we are then quotienting $T^6$ by the group with two generators,
\be\label{K3quotient}
g_1:  x \rightarrow -x, \qquad g_2: (y,z) \rightarrow (-y,-z),
\ee
where the geometric action generated by $g_1$ is combined with $\Omega (-1)^{F_L}$. We can view the background created by the quotient~\C{K3quotient}\ as a limit of the smooth case with only $O7$-planes and D7-branes. In this case, the tadpole is ${\chi(\M) \over 24} = 24$ units of brane charge. 

There is another way to view this orbifold limit as a background with $O3$-planes as well as $O7$-planes. The $O3$-planes are generated by the group element $g_1g_2$. In this case, there is a puzzle concerning the computation of the D3-brane tadpole. On the one hand, the $64$ $O3$-planes generate $-16$ units of D3-brane charge. On the other hand, we expect each $O7$-plane, together with its collection of $4$ D7-branes, to generate $-6$ units of D3-brane charge. A heuristic resolution of this puzzle leads to a quite interesting picture. In computing the Euler characteristic of $T^4/\Z_2$ for each of the $O7$-D7 systems, it appears we should only use the untwisted sector cohomology rather than the cohomology of the resolved space. In this case, the $\Z_2$ projects out all the odd cohomology giving
\be
\chi(T^4/\Z_2) = b_0 + b_2 + b_4 = 1 + 6 + 1 = 8.
\ee
Each $O7$-D7 system then contributes $-\chi(T^4/\Z_2)/4 = -2$ units of D3-brane charge. In total we find $-16$ units from the $O3$-planes and $-8$ units from the $O7$-D7 systems for a total of $-24$ units of D3-brane charge. 
We should stress that this method of counting should be applied with care to the perturbative orientifold locus. A perturbative orbifold action generates a $B$-field at each orbifold point. The $B$-field correlates with a gauge bundle of a topological type that admits no vector structure~\cite{Angelantonj:1996uy, Sen:1997pm}. A T-dual description of this particular model with D9 and D5-branes was nicely analyzed in~\cite{Berkooz:1996iz}, where gauge-bundle instantons provided $16$ units of D5-brane charge leaving a deficit of $8$ units of D5-brane charge.  The method of computing the D3-brane tadpole described above really applies to the geometric quotient. 

Now let us consider the choice of fluxes. Restricting to $4$-form fluxes compatible with the quotient action~\C{mquotient} requires the following additional constraints on the coefficients appearing in~\C{fluxparameters}:
\be
a_6=a_7= a_9=a_{10}=0.
\ee
Let us again set $e^{\phi_B}=1$. There are again many ways to solve the tadpole constraint. Let us first look at the simplest example with only $b_1\neq 0$. In this case, inspecting~\C{b1only}\ gives the quantization condition $2b_1\in \Z$. The tadpole constraint becomes, 
\be
4(2b_1)^2 + \# {\rm branes} = 24. 
\ee
Choosing $2b_1=2$ and adding $8$ D3-branes solves the tadpole constraint. 

If we want to solve the tadpole constraint without branes, we again need to turn on some $a$ coefficients in~\C{fluxparameters}. Let us return to the case~\C{a2a8}\ with $a_2$ and $a_8$ non-vanishing. Conveniently, both these fluxes are invariant under the quotient action.  We see that flux quantization requires $a_2\in\Z$ and $a_8\in\Z$. The tadpole constraint becomes,
\be
4(2b_1)^2 +8 (a_2)^2+ 8(a_8)^2 = 24. 
\ee 
Choosing $2b_1=2$ and $a_2=1, a_8=0$ solves the tadpole constraint with no brane sources. 

Now we dualize along $(x_1, y_1, z_1)$. The resulting orientifold action is generated by, 
\be\label{orientifoldaction}
{\hat g}_1:  (x_2, y_1, z_1) \rightarrow (-x_2, -y_1, -z_1), \qquad {\hat g}_2: (y,z) \rightarrow (-y,-z),
\ee
where the geometric action generated by ${\hat g}_1$ is again combined with $\Omega (-1)^{F_L}$. Each $O7$-plane becomes an $O6$-plane localized in the $(x_2, y_1, z_1)$ directions. Each $O3$-plane becomes an $O6$-plane localized in the $(x_2, y_2, z_2)$ directions. The resulting background only has $O6$-planes.

\begin{table} \begin{center}
\begin{tabular}{c|cccc|cccccc} 
& 0 & 1 & 2 & 3 & $x_1$ & $x_2$ & $y_1$ & $y_2$ & $z_1$ & $z_2$ \\
\hline
$O6$ & x & x & x & x & x &  &  & x &  & x \\
D6 & x & x & x & x & x &  & x  &  & x &  \\
\end{tabular} \vskip 0.5cm \caption{The depicted $O6$-planes are dual to the initial type IIB $O7$-planes. The D6-branes are dual to any initial type IIB D3-branes. The charge tadpole for these D6-branes is dual to the original type IIB D3-brane tadpole. That couplings on the $O6$-planes contribute to this D6-brane charge tadpole is the surprising phenomenon. } \label{figuretadpole} \end{center}\end{table}

The D7-branes of the original type IIB background become D6-branes transverse to the $(x_2, y_1, z_1)$ directions. These D6-branes cancel the charge of the $O6$-planes localized in the $(x_2, y_1, z_1)$ directions. This reflects the pointwise cancelation of D7-brane charge in the original IIB theory.  Any D3-branes present in type IIB also become D6-branes transverse  to the $(x_2, y_2, z_2)$ directions. These D6-branes {\it do not} cancel the charge of the $O6$-planes localized in the $(x_2, y_2, z_2)$ directions. We know there must be an additional contribution from flux sources and from the $O6$-planes in the $(x_2, y_1, z_1)$ directions.  We will take a closer look at tadpole cancelation in this massive IIA background in the following section. 

Again it is natural to suspect that one can remove the ${\hat g}_2$ quotient action appearing in~\C{orientifoldaction}\ by replacing the metric in the $(y,z)$ directions by a smooth $K3$ metric and suitably defining ${\hat g}_1$. In such a case, there would only be $O6$-planes and D6-branes localized in the $(x_2, y_1, z_1)$ directions, but they would produce a D6-brane tadpole for branes transverse to the $(x_2, y_2, z_2)$ directions, as in the toroidal orientifold case above. The orientation of these branes is depicted in table~\ref{figuretadpole}. The $O6$-planes and D6-branes are generating D6-brane charge in directions normal to their world-volumes.  This is a T-dual version of the phenomenon described in~\cite{McOrist:2012yc}, which is a central ingredient for evading supergravity no-go theorems in M-theory and type IIA string theory. It would be very interesting to explore this potentially smooth example.

\section{The Type II SUGRA Equations of Motion}\label{demfluxes}

Before examining specific examples in more detail, let us present the type II supergravity equations of motion in a formulation that treats all the R-R field strengths democratically. This approach has been taken in~\cite{Fukuma:1999jt, Bergshoeff:2001pv, Koerber:2010bx}. A subsequent more complete treatment of R-R fields in the framework of differential K-theory can be found in~\cite{Belov:2006jd, Belov:2006xj}. This formulation is very useful for checking equations of motion and Bianchi identities, especially when using T-duality. In this paradigm, we work with a complete set of R-R field strengths, either 
\be \{F_0,F_2,F_4,F_6,F_8,F_{10}\} \ee 
for IIA, or 
\be \{F_1,F_3,F_5,F_7,F_9\} \ee for IIB.  There are a number of places where sign conventions are rather important. We have described the places where sign ambiguities lurk in  Appendix~\ref{signconventions},  with our final choice of sign conventions stated in~\C{finalsigns}. We impose the duality relations,
\bea
&\ast F_0 = -F_{10}, \qquad \ast F_1 = F_9, \qquad \ast F_2 = F_8, & \label{dualityrelations1}\\
&\ast F_3 = -F_7, \qquad \ast F_4 = -F_6, \qquad \ast F_5 = F_5, & \label{dualityrelations2}\\
&\ast F_6 = F_4, \qquad \ast F_7 = -F_3, \qquad \ast F_8 = -F_2, & \label{dualityrelations3}\\
&\ast F_9 = F_1, \qquad \ast F_{10} = F_0, &  \label{dualityrelations4}
\eea
by hand at the level of the equation of motion.  This puts all the R-R fields on the same footing as $F_5$, and also makes it easy to talk about either IIA or IIB at the same time.  It also has the advantage of simplifying the R-R part of the bosonic action (when written in terms of field strengths), because the Chern-Simons terms are absorbed into the kinetic terms.  

For either IIA or IIB, we have the fairly simple action:
\be
S_{II}=\frac{1}{2\k^2}\int d^{10}x\sqrt{-g}\left\{ e^{-2\Phi}\lp R+4\left|d\Phi\right|^2-\hlf\left|H\right|^2\rp-\frac{1}{4}\sum_p\left|F_p\right|^2\right\}.
\ee
How do we make use of this simple form?  To begin, we note that the Bianchi identity
\be
\label{maintextBianchi}
dF_p=- H\w F_{p-2},
\ee
implies that each field strength can be written locally in the form
\be
\label{eq:RRPotentials}
F_p=e^{-B}\lp dC+m\rp|_{p\mathrm{-form}},
\ee
where $C=\sum_pC_p$, and we have included the possibility of a constant $m$ in IIA. This constant corresponds to the Romans mass.  Note that the potentials defined in this way will be inconvenient for many purposes; for example, they will have ugly transformations under NS-NS gauge transformations. In addition, the action expressed in terms of these potentials will not be pretty. However, they will be related to other more convenient choices like~\C{rrfieldstrengths}\ by field redefinitions, and this particular choice will help us derive the equations of motion.

We claim that the R-R equations of motion will be satisfied provided we satisfy the duality relations~\C{dualityrelations1},~\C{dualityrelations2},~\C{dualityrelations3}, and~\C{dualityrelations4}, and we check that the Bianchi identities~\C{maintextBianchi}\ are satisfied.  Let us see how this works in IIA.  
Consider $C_9$ which only appears in the action via the term,
\be
-\frac{1}{4}\left|dC_9-B\w dC_7+\ldots\right|^2.
\ee
The corresponding equation of motion is simply,
\be
d\ast F_{10}=0.
\ee
Now that we have an equation of motion, we can substitute in the duality relation $\ast F_{10}=F_0$, and the equation becomes $dF_0=0$, which is simply the Bianchi identity for $F_0$.

Next we can consider $C_7$ which appears in two terms, leading to the equation of motion:
\be
d\lp\ast F_8-B\w\ast F_{10}\rp=0.
\ee
Using the duality relations, this becomes
\be
0=-d\lp F_2+BF_0\rp=-\lp\lp dF_2+HF_0\rp+B\w\lp dF_0\rp\rp,
\ee
which holds by the Bianchi identities for $F_2$ and $F_0$.

Let us check one more case; $C_5$ leads to an equation of motion, 
\bea
0&=& d\lp\ast F_6-B\w\ast F_8+\hlf B^2\w\ast F_{10}\rp=d\lp F_4+B\w F_2+\hlf B^2F_0\rp\cr
&=& \lp\lp dF_4+H\w F_2\rp+B\w\lp dF_2+HF_0\rp+\hlf B^2\w\lp dF_0\rp\rp.
\eea
Hopefully this is convincing evidence for the claim that the equations of motion are equivalent to the duality relations plus Bianchi identities.

From (\ref{eq:RRPotentials}), we also have the result that varying $F_p$ with respect to $B$ gives us $-F_{p-2}$, so the equation of motion for $B$ becomes simply
\be
0=d\lp e^{-2\Phi}\ast H\rp+\hlf\sum_{p\ge 2}F_{p-2}\w\ast F_p.
\ee
The dilaton equation of motion is also quite simple,
\be
R+4\nabla^2\Phi-4\lp\nabla\Phi\rp^2-\hlf\left|H\right|^2=0.
\ee

This leaves only the Einstein equations. After subtracting a multiple of the dilaton equation, these equations become, 
\begin{multline}
e^{-2\Phi}\lp R_{\m\n}+2\nabla_\m\nabla_\n\Phi-\frac{1}{4}H_\m^{\hph{\m}\rho\s}H_{\n\rho\s}\rp\\
+\frac{1}{8}g_{\m\n}\sum_p\left|F_p\right|^2-\frac{1}{4(p-1)!}\sum_{p\ge 1}F_{p\,\m}^{\hph{p\,\m}\rho_1\cdots\rho_{p-1}}F_{p\,\n\rho_1\cdots\rho_{p-1}}=0.
\end{multline}
There is even a further simplification; once we impose self-duality on the R-R field strengths then $\sum_p|F_p|^2=0$.  For example, $|F_5|^2$ is proportional to $\ast(F_5\w\ast F_5)=\ast(F_5\w F_5)=0$, and $|F_4|^2+|F_6|^2$ is proportional to $\ast(F_4\w (-F_6)+F_6\w F_4)=0$.

To summarize, the type II equations of motion are:
\bea
R+4\nabla^2\Phi-4\lp\nabla\Phi\rp^2-\hlf\left|H\right|^2 &=& 0\qquad\mathrm{(dilaton)},\\
e^{-2\Phi}\lp R_{\m\n}+2\nabla_\m\nabla_\n\Phi-\hlf\left|H\right|^2_{\m\n}\rp-\frac{1}{4}\sum_{p\ge 1}\left|F_p\right|^2_{\m\n} &=& 0\qquad\mathrm{(Einstein)},\\
d\lp e^{-2\Phi}\ast H\rp+\hlf\sum_{p\ge 2}F_{p-2}\w\ast F_p &=& 0\qquad\mathrm{(}B\mathrm{-field)},\\
dF_p+H\w F_{p-2} &=& 0\qquad\mathrm{(Bianchi)},\\
\ast F_p+\lp -1\rp^{p(p+1)/2} F_{10-p} &=& 0\qquad\mathrm{(duality)},
\eea
where in the Einstein equations we defined:
\be
\left|\om^{(p)}\right|^2_{\m\n}=\frac{1}{(p-1)!}\om_\m^{\hph{\m}\rho_1\cdots\rho_{p-1}}\om_{\n\rho_1\cdots\rho_{p-1}}.
\ee

\section{A Closer Look at Specific Examples}\label{simple}

We want to first examine some of the general features of these models. Let us again restrict to  T-duality in the $(x_1, y_1, z_1)$ direction with the general T-dual model described in section~\ref{generalcase}. For this general case, we find models in both conventional type IIA string theory and in massive IIA when the $b_1$ parameter of~\C{fluxparameters}\ is non-zero. Performing three T-dualities is a mirror transform in the absence of flux. Some local aspects of the mirror transform with fluxes have been discussed in~\cite{Fidanza:2003zi, Grana:2006hr}. 

The original type IIB flux vacuum has a number of K\"ahler moduli, including the total volume of the space. 
We can view the total volume as the modulus corresponding to dilating the internal warp factor of~\C{iibmetric}\ by sending 
\be\label{dilation} w\rightarrow w+\lambda \ee 
for some positive $\lambda$. The internal  metric in string-frame for the dual theory takes the form, 
\bea\label{dualmetric}
ds_{dual}^2 &&=  e^{w+ {\phi_B\over 2}} \left\{ (dx_2)^2  +  (dy_2)^2 +  (dz_2)^2 \right\}  \cr && {\phantom =} + e^{-w- {\phi_B\over 2}} \left\{ (dx_1+A_x)^2  +  (dy_1+A_y)^2 +  (dz_1+A_z)^2 \right\}. 
\eea
For the simplest case where $A_i=0$, the overall type IIB volume scaling contracts the $(x_1, y_1, z_1)$ $T^3$-factor while dilating the $(x_2, y_2, z_2)$ $T^3$-factor exactly as we would expect after applying $3$ T-dualities. Rather the total volume of the space is set by the complex structure of the type IIB model, with the complex structure typically completely fixed by the choice of flux. 

The background~\C{dualmetric}\ is not a large volume compactification, and we should expect large $\alpha'$ corrections to the tree-level physics. This is precisely the same situation seen in the torsional heterotic/type I models found by a similar procedure. Regardless of whether there are large corrections, the form of the solutions is rather interesting. In the torsional setting, those models provide examples of the topological type of compact non-K\"ahler spaces that solve the heterotic/type I Bianchi identity. We expect something similar here for type IIA and massive IIA.  

Models in type IIA with $F_0=0$ can be supersymmetric, preserving at least $N=1$ and $N=2$ space-time supersymmetry. More exotic choices like $N=3$ might also be possible in this setting~\cite{Frey:2002hf}. However, all the models we have described involve only $F_2$-flux and $O6$-planes together with D6-branes. With $F_0=0$, there is no $H_3$-flux or $F_4$-flux. Therefore each such model should be liftable to a geometric background of M-theory with a metric that preserves the appropriate amount of supersymmetry.  The CY geometries found in the lift  to M-theory of models preserving $N=2$ supersymmetry have been studied for specific cases in~\cite{Sethi:1996es, Schulz:2004tt}. If the lifts of $N=1$ models are classical backgrounds then they will be $G_2$ holonomy geometries that have yet to be described. However, they need not be classical! Rather, one might suspect that these backgrounds solve the gravity equations of motion only when $R^4$ corrections are included in the general case. This would be quite fascinating to explore further.  

However, our main emphasis in this work are models with a Romans mass. In this case, there is an $H_3$-flux precisely correlated with $F_0$ so these are honest flux backgrounds. It would be nice if there were a way to geometrize the $F_2$-flux data in the Romans theory, as in the M-theory lifts of type IIA vacua with only $F_2$-flux. Perhaps some extension of the attempt to geometrize massive IIA found in~\cite{Hull:1998vy}\ might be useful.

 Solving the massive IIA equations of motion with scale separation is highly non-trivial. We know of no other examples with either Minkowski or $AdS_4$ space-times  with scale separation so it seems worthwhile to explore how these solutions work. The  issue to which we first turn is the manner in which the D6-brane charge Bianchi identity is being satisfied.  

\subsection{The Bianchi identity for D6-brane charge}

Under three T-dualities, the original type IIB D3-brane tadpole becomes a D6-brane charge tadpole for D6-branes localized in the $(x_2,y_2,z_2)$ directions. The D3-brane tadpole is a central ingredient in evading the usual no-go theorem forbidding supergravity flux compactifications. To see how this tadpole condition is satisfied in type IIA, let us examine the Bianchi identity for $F_2$. 

Using the definitions in~\C{massivefieldstrengths}, the Bianchi identity for the massive IIA $2$-form field strength is
\be\label{F2bianchi}
dF_2 = - mH_3=- F_0H_3,
\ee
where $F_2$ and $F_0$ are the general fluxes appearing in~\C{generalfluxes}. In addition to the supergravity source appearing on the right hand side, there will be contributions to~\C{F2bianchi}\ from $O6$-planes and D6-branes, including the unusual contributions from higher derivative interactions supported on $O6$-planes described in section~\ref{oplanes}.  

In the general case of section~\C{generalcase}, the warp factor equation following from~\C{warp}\ is
\be\label{generalwarp}
\delta^{ab}\partial_a\partial_b e^{2w} = - 32 e^{-\phi_B} \left[ 2 (b_1)^2 + a_2a_3 + a_2 a_4 +a_3 a_4 +\sum_{i=2}^{10} (a_i)^2\right],
\ee
with $a,b$ labeling the coordinates of the $(x_2,y_2,z_2)$ torus.
For transparency, let us consider a less-than-general case with only the $a_6 = a_{10}$ and $b_1$ fluxes turned on. Then,
\be
F_2 = -8a_6e^{-\phi_B}dx_2 \wedge \lp dz_1 + A_z \rp + \ast_3 d_3 \lp e^{2w} \rp.
\ee
The exterior derivative of this expression gives,
\bea
dF_2 &&= 8 a_6 e^{-\phi_B} dx_2 \wedge dA_z + d \lp \ast_3 d_3 \lp e^{2w} \rp \rp \\ \non
&& = 64 a_6^2 e^{-\phi_B} dx_2 \wedge dy_2 \wedge dz_2 +  d\lp \ast_3 d_3 \lp e^{2w} \rp \rp .
\eea
Combining this with the right hand side of~\C{F2bianchi} gives us back the desired specific terms in the warp factor equation~\C{generalwarp}, 
\be
64e^{-\phi_B} \lp a_6^2 + b_1^2 \rp dx_2\wedge dy_2 \wedge dz_2 = -  d \lp \ast_3 d_3 \lp e^{2w} \rp \rp ,
\ee
provided that the warp factor only depends on the coordinates of the three-torus given by $(x_2,y_2,z_2)$.
This assumption is perfectly fine for checking that the D3-brane tadpole maps to the D6-brane tadpole of the dual theory. For example, we could have considered a non-compact example with $(x_2,y_2,z_2)\in \R^3$ and a $T^3$-fiber with coordinates $(x_1, y_1, z_1)$. In this case, the warp factor equation can be solved in supergravity with only dependence on  $(x_2,y_2,z_2)$ and no exotic additional sources. If we allowed a more general choice of $a_i$ and $b_i$ coefficients, we would find equation~\C{generalwarp}\ reproduced by the $F_2$ Bianchi identity. 

We can now see why there is a precise correlation between $H_3$ and $F_0$. In the original type IIB frame, there is a tadpole contribution proportional to $(b_1)^2$ precisely from the fluxes that dualize to these terms, which is reproduced by the right hand side of~\C{F2bianchi}. On the other hand, the source of the $a_i$ terms in the type IIA tadpole condition is the manifest non-closure of $F_2$.

\subsection{The equations of motion}

The tadpole condition is the most basic check. A much stronger constraint is satisfying the equations of motion. We will check the equations of motion are satisfied for the simplest example of section~\ref{simplest}. There are remarkably few explicit checks of equations of motion for flux backgrounds. We find it gratifying that these metrics and fluxes do indeed solve the equations of motion, modulo the inclusion of beyond supergravity sources needed for compact solutions.  

\subsubsection{type IIB starting point -- the simplest case}

Our starting point is IIB string theory with constant dilaton, $\phi_B$, and metric
\be\label{startmetric}
ds^2=e^{-w+\phi_B/2}\eta_{\m\n}dx^\m dx^\n+e^{w+\phi_B/2}\d_{AB}dx^Adx^B,
\ee
where $\m,\n=0,\ldots, 3$ and $A,B=4,\ldots,9$. The warp factor, $w$, depends on $x^A$ only. We will later split these coordinates further into,
\be
\left\{x^4,\cdots,x^9\right\}=\left\{x_1,x_2,y_1,y_2,z_1,z_2\right\},
\ee
and restrict $w$ to dependence only on $\{x_2,y_2,z_2\}$, but there is no reason do that yet.  The metric~\C{startmetric}\ has determinant,
\be
\sqrt{-g}=e^{w+5\phi_B/2},
\ee
and Ricci tensor,
\be
R_{\m\n}=\hlf\eta_{\m\n}e^{-2w}\d^{AB}\p_A\p_Bw,\qquad R_{AB}=-\hlf\d_{AB}\d^{CD}\p_C\p_Dw-2\p_Aw\p_Bw.
\ee
Thus the Ricci scalar is
\be
R=-e^{-w-\phi_B/2}\d^{AB}\lp\p_A\p_Bw+2\p_Aw\p_Bw\rp=-\hlf e^{-3w-\phi_B/2}\d^{AB}\p_A\p_B\lp e^{2w}\rp.
\ee
We will also turn on fluxes
\be
H_3=-8 b_1dx_2\w dy_2\w dz_2,\qquad F_3=8 b_1e^{-\phi_B}dx_1\w dy_1\w dz_1,
\ee
and
\be
F_5=-\lp 1+\ast\rp\lp d\lp e^{-2w}\rp\w\vol_4\rp.
\ee
By~\C{dualityrelations2}, we have also
\be
F_7=8b_1e^{-2w}\vol_4\w dx_2\w dy_2\w dz_2.
\ee

By construction the R-R fluxes are properly self-dual, so we need only check the Bianchi identities.  For $F_3$, we just have $dF_3=0$, which is satisfied since $F_3$ is closed.  For $F_7$, we require 
\bea
 dF_7+H\w F_5  &=& 8b_1d\lp e^{-2w}\rp\w\vol_4\w dx_2\w dy_2\w dz_2  \\ && +
\lp 8 b_1dx_2\w dy_2\w dz_2\rp\w\lp d\lp e^{-2w}\rp\w\vol_4\rp \cr &=& 0, \non 
\eea
which is true since we have been very careful with signs.

Next we turn to the Bianchi equation for $F_5$,
\bea 
0&=& dF_5+H\w F_3\\
&=& -d\ast\ls d\lp e^{-2w}\rp\w\vol_4\rs-64b_1^2e^{-\phi_B}dx_2\w dy_2\w dz_2\w dx_1\w dy_1\w dz_1.\non
\eea
Since we can compute
\be
d\ast\ls d\lp e^{-2w}\rp\w\vol_4\rs=\d^{AB}\p_A\p_B\lp e^{2w}\rp dx_2\w dy_2\w dz_2\w dx_1\w dy_1\w dz_1,
\ee
we find that this reduces to a Poisson equation for the warp factor,
\be\label{eq:T6WarpFactorEq}
\d^{AB}\p_A\p_B\lp e^{2w}\rp=-64 b_1^2e^{-\phi_B}.
\ee
We note that this is only a local equation.  On a compact space we will not be able to solve it globally unless there are extra source contributions; for example, from orientifold planes, D-branes and their supported higher derivative couplings.

Let's turn now to the NS-NS sector equations.  The dilaton equation is simply
\be
0=R-\hlf\left|H\right|^2=-\hlf e^{-3w-\phi_B/2}\d^{AB}\p_A\p_B\lp e^{2w}\rp-32b_1^2e^{-3w-3\phi_B/2},
\ee
and it is easy to see that this is equivalent to (\ref{eq:T6WarpFactorEq}).

The equation from varying the $B$-field becomes,
\bea
d\lp e^{-2\Phi}\ast H\rp+\hlf F_3\w\ast F_5+\hlf F_5\w\ast F_7&=& d\lp e^{-2\Phi}\ast H\rp+F_3\w F_5 \\
&=& d\lp -8b_1e^{-2w-\phi_B}\vol_4\w dx_1\w dy_1\w dz_1\rp\cr
&& -\lp 8b_1e^{-\phi_B}dx_1\w dy_1\w dz_1\rp\w\lp d\lp e^{-2w}\rp\w\vol_4\rp \cr
&=&0. \non
\eea

Finally, we have the Einstein equations.  For the $\m-\n$ components, we find
\bea
0 &=& e^{-2\Phi}R_{\m\n}-\frac{1}{4}\lp F_5\rp^2_{\m\n}-\frac{1}{4}\lp F_7\rp^2_{\m\n}, \\
&=& \hlf\eta_{\m\n}e^{-2w-2\phi_B}\d^{AB}\p_A\p_Bw+\eta_{\m\n}e^{-2w-2\phi_B}\d^{AB}\p_Aw\p_Bw+16\eta_{\m\n}b_1^2e^{-4w-3\phi_B}, \non
\eea
which is again equivalent to~(\ref{eq:T6WarpFactorEq}).  The $A-B$ components give,
\be
0=e^{-2\Phi}\lp R_{AB}-\hlf\left|H\right|^2_{AB}\rp-\frac{1}{4}\sum_p\left| F_p\right|^2_{AB}.
\ee
Noting that we can rewrite
\be
-\hlf e^{-2\Phi}\left|H\right|^2_{AB}-\frac{1}{4}\left|F_3\right|^2_{AB}-\frac{1}{4}\left|F_7\right|^2_{AB}=-16b_1^2e^{-2w-3\phi_B}\d_{AB},
\ee
this becomes
\bea
0 &=& e^{-2\phi_B}\lp -\hlf\d_{AB}\d^{CD}\p_C\p_Dw-2\p_Aw\p_Bw\rp-16b_1^2e^{-2w-3\phi_B}\d_{AB}\\
&& +e^{-2\phi_B}\p_Aw\p_Bw-e^{-2\phi_B}\lp\d_{AB}\d^{CD}\p_Cw\p_Dw-\p_Aw\p_Bw\rp \cr
&=& e^{-2w-2\phi_B}\d_{AB}\lp -\frac{1}{4}\d^{CD}\p_C\p_D\lp e^{2w}\rp-16b_1^2e^{-\phi_B}\rp,
\eea
which is again equivalent to our warp factor equation. That concludes our check of the type IIB equations of motion. The democratic treatment of R-R fluxes, given in section~\ref{demfluxes}, greatly simplified this analysis.

\subsubsection{type IIA -- the simplest case}

After T-dualizing $x_1$, $y_1$, and $z_1$ in that order, we find a dilaton
\be
\phi_A=-\frac{3}{2}w+\frac{1}{4}\phi_B,
\ee
and a metric
\be
ds^2=e^{-w+\phi_B/2}\eta_{\m\n}dx^\m dx^\n+e^{w+\phi_B/2}\d_{ab}dx^adx^b+e^{-w-\phi_B/2}\d_{ij}dx^idx^j,
\ee
where $\m,\n=0,\ldots,3$, $x^a\in\{x_2,y_2,z_2\}$ and $x^i\in\{x_1,y_1,z_1\}$.  If the warp factor $w(x^a)$ is a function of $\{x_2,y_2,z_2\}$ then this metric has determinant,
\be
\sqrt{-g}=e^{-2w+\phi_B},
\ee
and a Ricci tensor given by,
\bea
R_{\m\n} &=& \hlf e^{-2w}\eta_{\m\n}\d^{ab}\lp\p_a\p_bw-3\p_aw\p_bw\rp,\\
R_{ab} &=& -\hlf\d_{ab}\d^{cd}\p_c\p_dw+3\p_a\p_bw+\frac{3}{2}\d_{ab}\d^{cd}\p_cw\p_dw-5\p_aw\p_bw,\\
R_{ij} &=& \hlf e^{-2w-\phi_B}\d_{ij}\d^{ab}\lp\p_a\p_bw-3\p_aw\p_bw\rp,
\eea
with Ricci scalar
\be
R=e^{-w-\phi_B/2}\d^{ab}\lp 5\p_a\p_bw-11\p_aw\p_bw\rp.
\ee
The $H$-flux is unchanged,
\be
H=-8b_1dx_2\w dy_2\w dz_2,
\ee
and the R-R fields become
\bea
&F_0=-8b_1e^{-\phi_B},\qquad F_{10}=8b_1e^{-2w}\vol_4\w dx_2\w dy_2\w dz_2\w dx_1\w dy_1\w dz_1, & \cr
&F_2=\ast_3d\lp e^{2w}\rp,\qquad F_8=-\vol_4\w d\lp e^{-2w}\rp\w dx_1\w dy_1\w dz_1, &
\eea
where $\ast_3$ represents the Hodge star operator with respect to the unwarped flat $T^3$ with coordinates $\{x_2,y_2,z_2\}$, using $\e_{x_2y_2z_2}=1$.

We can easily check that these R-R field strengths obey the correct duality conditions.  Bianchi is of course automatic for $F_0$ and $F_{10}$.  For $F_2$, we compute
\bea
dF_2+H_3F_0&=&\d^{ab}\p_a\p_b\lp e^{2w}\rp dx_2\w dy_2\w dz_2 +64b_1^2e^{-\phi_B}dx_2\w dy_2\w dz_2 \cr &=&0,
\eea
provided that the same warp factor equation we found before is satisfied,  
\be
\d^{ab}\p_a\p_b\lp e^{2w}\rp+64b_1^2e^{-\phi_B}=0,
\ee
where we are explicit about only allowing dependence on $\{x_2,y_2,z_2\}$. This latter restriction will be relaxed in the full solution with beyond supergravity sources. 
Finally, the Bianchi identity  for $F_8$ is satisfied because $dF_8=0$ and $F_6=0$.

Turning to the dilaton equation of motion, we need the observation that there are non-vanishing Christoffel symbols for this metric, leading to
\bea
\nabla_\m\nabla_\n\Phi &=& \frac{3}{4}e^{-2w}\eta_{\m\n}\d^{ab}\p_aw\p_bw,\\
\nabla_a\nabla_b\Phi &=& -\frac{3}{2}\p_a\p_bw+\frac{3}{2}\p_aw\p_bw-\frac{3}{4}\d_{ab}\d^{cd}\p_cw\p_dw,\\
\nabla_i\nabla_j\Phi &=& \frac{3}{4}e^{-2w-\phi_B}\d_{ij}\d^{ab}\p_aw\p_bw,
\eea
and
\be
\nabla^2\Phi=\frac{3}{2}e^{-w-\phi_B/2}\d^{ab}\lp -\p_a\p_bw+3\p_aw\p_bw\rp.
\ee
Thus the dilaton equation of motion becomes,
\bea
R+4\nabla^2\Phi-4\lp\nabla\Phi\rp^2-\hlf\left|H\right|^2 &=& e^{-w-\phi_B/2}\d^{ab}\lp -\p_a\p_bw-2\p_aw\p_bw\rp \cr && -32b_1^2e^{-3w-3\phi_B/2} \\
&=& 0,\non
\eea
which is again equivalent to the warp factor equation.

The $B$-field equation gives,
\bea
0 &=& d\lp e^{-2\Phi}\ast H\rp+\hlf F_0\ast F_2+\hlf F_8\ast F_{10}=d\lp e^{3w-\phi_B/2}\ast H\rp+F_0F_8\cr
&=& d\lp -8b_1e^{-2w-\phi_B}\vol_4\w dx_1\w dy_1\w dz_1\rp  \cr && +8b_1e^{-\phi_B}\vol_4\w d\lp e^{-2w}\rp\w dx_1\w dy_1\w dz_1,
\eea
which is satisfied.

We are only left with a check of the Einstein equations.  We find
\bea
&& e^{-2\Phi}\lp R_{\m\n}+2\nabla_\m\nabla_\n\Phi\rp-\frac{1}{4}\left|F_8\right|^2_{\m\n}-\frac{1}{4}\left|F_{10}\right|^2_{\m\n}\cr
&& =\hlf e^{w-\phi_B/2}\eta_{\m\n}\d^{ab}\p_a\p_bw+e^{w-\phi_B/2}\eta_{\m\n}\d^{ab}\p_aw\p_bw+16b_1^2e^{-w-3\phi_B/2} \\ &&=0, \non
\eea
using the warp factor equation again.  The $ij$ component is of identical form.  Lastly we have
\bea
&& e^{-2\Phi}\lp R_{ab}+2\nabla_a\nabla_b\Phi-\hlf\left|H\right|^2_{ab}\rp-\frac{1}{4}\left|F_2\right|^2_{ab}-\frac{1}{4}\left|F_8\right|^2_{ab}-\frac{1}{4}\left|F_{10}\right|^2_{ab}\cr
&& {\phantom  {e^{-2\Phi}}}=e^{3w-\phi_B/2}\lp -\hlf\d_{ab}\d^{cd}\p_c\p_dw-2\p_aw\p_bw-32b_1^2e^{-2w-\phi_B/2}\d_{ab}\rp\cr
&& {\phantom {e^{-2\Phi}=}}-e^{3w-\phi_B/2}\lp\d_{ab}\d^{cd}\p_cw\p_dw-\p_aw\p_bw\rp+e^{3w-\phi_B/2}\p_aw\p_bw+16b_1^2e^{w-3\phi_B/2}\cr
&& {\phantom  {e^{-2\Phi}}}=e^{w-\phi_B/2}\d_{ab}\ls -\frac{1}{4}\d^{cd}\p_c\p_d\lp e^{2w}\rp-16b_1^2e^{-\phi_B}\rs,
\eea
and the warp factor equation appears one last time, so everything is consistent.

\section*{Acknowledgements}

We would like to thank Callum Quigley for participating in the early stages of this project.
The massive IIA examples are dedicated to Alessandro Tomasiello, who requested an example of a massive IIA flux vacuum with scale separation and no smeared orientifolds. It is our pleasure to thank Mariana Grana, Chris Hull and Michael Schulz for helpful comments.  T.~M. and S.~S. are supported in part by
NSF Grant No.~PHY-1316960.  D.~R. is supported by funding from the European Research Council, ERC Grant agreement no.\ 268088-EMERGRAV.

\newpage
\appendix

\section{Sign Manifesto}\label{signconventions}

We need to be quite careful about signs so we will keep track of the possible sign choices by allowing $\al_i=\pm$ to appear in various locations.  In the main text, we will use a specific choice for $\al_i$ stated below.  For example, we take the convention that the epsilon tensors (not symbols) satisfy:
\be
\e_{01\cdots 9}=\al_1\sqrt{-g^{(10)}},\qquad\e_{0123}=\al_1\sqrt{-g^{(4)}}.
\ee
The factors of the square root of the metric determinant ensure that $\e$ transforms covariantly.

We define the Hodge star as follows,
\be
\ast\lp dx^{\m_1}\w\cdots\w dx^{\m_p}\rp=\frac{1}{(d-p)!}\e^{\m_1\cdots\m_p}_{\hph{\m_1\cdots\m_p}\n_1\cdots\n_{d-p}}dx^{\n_1}\w\cdots\w dx^{\n_{d-p}}.
\ee
In components, this reads:
\be
\lp\ast\om^{(p)}\rp_{\n_1\cdots\n_{d-p}}=\frac{1}{p!}\e^{\m_1\cdots\m_p}_{\hph{\m_1\cdots\m_p}\n_1\cdots\n_{d-p}}\om^{(p)}_{\m_1\cdots\m_p}.
\ee
Note that for $d$ even and $p$ odd (such as a five-form in ten dimensions), this differs by a sign from the convention found in Polchinski (B.4.6)~\cite{Polchinski:1998rr}.  Because this might cause issues, we will also insert a sign into the duality relation for $F_5$; namely our $F_5$ will obey,
\be
\ast F_5=\al_2F_5,
\ee
in ten dimensions. With our choice of conventions for the T-duality rules given in Appendix~\ref{tduality}, this choice implies that:\footnote{This can be seen as follows: for $10$-dimensional flat space, a valid $F_5$ in our language would be $dx^{01234}-\al_1\al_2dx^{56789}$.  Repeated T-dualities along the $(5, \ldots, 9)$ directions give the stated  relations.}
\bea
\label{eq:starF0}
\ast F_0 &=& -\al_2F_{10},\\
\ast F_1 &=& \al_2F_9,\\
\ast F_2 &=& \al_2F_8,\\
\label{eq:starF3}
\ast F_3 &=& -\al_2F_7,\\
\ast F_4 &=& -\al_2F_6,\\
\ast F_5 &=& \al_2F_5,\\
\ast F_6 &=& \al_2F_4,\\
\ast F_7 &=& -\al_2F_3,\\
\ast F_8 &=& -\al_2F_2,\\
\ast F_9 &=& \al_2F_1,\\
\label{eq:starF10}
\ast F_{10} &=& \al_2F_0.
\eea
Let us also insert a sign in the Bianchi identity,
\be
\label{eq:Bianchi}
dF_p=-\al_3H\w F_{p-2},
\ee
and assume that this relation will hold with the same sign in both IIB and (massive) IIA. We use the mostly plus convention for the metric signature in Minkowski space, as do all right-hearted people.

 A useful relation for us follows from defining,
\be
\left|\om^{(p)}\right|^2=\frac{1}{p!}\om^{\m_1\cdots\m_p}\om_{\m_1\cdots\m_p},
\ee
from which we find,
\be
\int d^dx\sqrt{-g}\left|\om^{(p)}\right|^2=-\al_1\int\om_p\w\ast\om_p.
\ee

In passing let us note that as a result of these conventions, the volume form in flat $\R^{1,3}$ is
\be\label{volumefour}
 \vol_4=\sqrt{-g}dx^0\w dx^1\w dx^2\w dx^3=\frac{\al_1}{4!}\e_{\m_1\m_2\m_3\m_4}dx^{\m_1}\w dx^{\m_2}\w dx^{\m_3}\w dx^{\m_4},
\ee
and satisfies
\be
\ast \vol_4=\sqrt{-g}\e^{0123}=-\al_1,\qquad \ast 1=\al_1 \vol_4.
\ee
Rather than clutter the main text by leaving these $\alpha_i$ arbitrary, we will make a specific choice:
\be\label{finalsigns}
\al_i=1. 
\ee
This also determines the various sign choices in the explicit expressions for flux appearing in the main text. 

\section{T-duality Rules}\label{tduality}

For a string background specified by a metric $g$, $B$-field and dilaton $\phi$ with isometry in the $y$-direction, T-duality applied to the background gives a new background in terms of primed fields,
\bea
g'_{yy} = {1\over g_{yy}}, \qquad g'_{\mu y} = {B_{\mu y} \over g_{yy}}, \qquad g'_{\mu\nu} = g_{\mu\nu} - {g_{\mu y} g_{\nu y} - B_{\mu y} B_{\nu y} \over g_{yy}}, \\
B'_{\mu y} = {g_{\mu y} \over g_{yy}}, \qquad B'_{\mu\nu} = B_{\mu\nu} - {B_{\mu y} g_{\nu y} - g_{\mu y} B_{\nu y}\over g_{yy}}, \qquad \phi' = \phi - {1\over 2} {\rm log}( g_{yy}).
\eea
The definition of the R-R field strengths~\C{eq:RRPotentials}\ in terms of potentials given in the main text is a particularly convenient trivialization for deriving the equations of motion. An alternate trivialization, which often appears in discussions of T-duality in the literature, defines potentials via:  
\be\label{rrfieldstrengths}
F^{(p)} = dC^{(p-1)} + H\wedge C^{(p-3)}.
\ee 
The two definitions~\C{rrfieldstrengths}\ and~\C{eq:RRPotentials}\  are related by a field redefinition. However, the T-duality transformations given below do not depend on the choice of trivialization. Rather, they depend only on the equations of motion and Bianchi identities. 


In the main text, we use either  $F^{(p)}$ or $F_p$ to denote the R-R field strengths. In this appendix, it is useful to use $F^{(p)}$ to avoid index confusion.  For the most part, we will only be concerned with the field strengths themselves. The R-R potentials are implicitly defined by~\C{rrfieldstrengths}. T-duality acts on these field strengths as follows~\cite{Bergshoeff:1996ui, Bergshoeff:1996cy, Green:1996bh, Hassan:1999bv, Fukuma:1999jt}, 
\bea
&& F^{(p)'}_{\mu_1 \cdots \mu_{p-1} y}  = F^{(p-1)}_{\mu_1 \cdots \mu_{p-1}} - (p-1) { F^{(p-1)}_{[\mu_1 \cdots \mu_{p-2} |y|} g_{\mu_{p-1}]y} \over g_{yy}}, \\
&& F^{(p)'}_{\mu_1 \cdots \mu_{p}} = F^{(p+1)}_{\mu_1 \cdots \mu_{p} y} + p F^{(p-1)}_{[\mu_1 \cdots \mu_{p-1}} B_{\mu_p]y}+p(p-1) {F^{(p-1)}_{[\mu_1\cdots\mu_{p-2}|y|}B_{\mu_{p-1}|y|}g_{\mu_p]y} \over g_{yy}}.
\eea
The virtue of using this definition is the ability to dualize backgrounds for which the R-R field strengths respect the chosen isometries, while the associated R-R potentials need not. Dualizing to massive IIA always involves a situation of this type. Note that $F^{(5)}$ given by~\C{rrfieldstrengths}\  obeys self-duality,
\be
F^{(5)} = \ast F^{(5)}, 
\ee
in accord with~\C{finalsigns}. This equation gives rise to the warp factor equation~\C{warp}. We can extend the definition~\C{rrfieldstrengths}\ to massive IIA with mass parameter $m$ by defining:
\be\label{massivefieldstrengths}
F^{(0)} = m, \qquad  F^{(2)}= dC^{(1)} - m B, \qquad F^{(4)}= dC^{(3)} + H\wedge C^{(1)} + {1\over 2} m B\wedge B.
\ee

\newpage

\ifx\undefined\bysame
\newcommand{\bysame}{\leavevmode\hbox to3em{\hrulefill}\,}
\fi

\end{document}